\documentclass[aps,pra,twocolumn,twoside,floatfix,a4,showpacs,superscriptaddress]{revtex4}

\usepackage{float}
\usepackage{caption}
\usepackage{bbm}
\usepackage{amsmath, amssymb}
\usepackage{color}
\usepackage{graphicx,epsfig}
\usepackage{times} 
\usepackage{braket}
\usepackage{amsfonts}
\usepackage{amsthm}
\usepackage{amsmath}
\usepackage{multirow}
\usepackage[mathscr]{euscript}
\usepackage[stable]{footmisc}
\usepackage{MnSymbol}

\newfloat{graphic}{tbp}{lgr}   
\floatname{graphic}{Graphic}

\newcommand{\hus}{\textcolor{black}}

 \newcommand{\earl}[1]{#1}  
 
\renewcommand\vec[1]{\ensuremath\boldsymbol{#1}}
\newcommand{\tr}{\ensuremath{\mathrm{tr}}}

\newcommand{\kb}[2]{\ensuremath{\vert #1 \rangle \langle #2 \vert}}

\begin{document}

\title{\textbf{Qutrit Magic State Distillation}}
\author{Hussain Anwar}
\email{hussain.anwar.09@ucl.ac.uk}
\affiliation{Department of Physics and Astronomy, University College London, Gower Street, London WC1E 6BT, United Kingdom.}
\author{Earl T. Campbell}
\affiliation{\textit{Dahlem Center for Complex Quantum Systems, Freie Universitat Berlin, 14195 Berlin, Germany.}}
\author{Dan E. Browne}
\affiliation{Department of Physics and Astronomy, University College London, Gower Street, London WC1E 6BT, United Kingdom.}
  
\begin{abstract}
Magic state distillation (MSD) is a purification protocol that plays a central role in fault tolerant quantum computation. Repeated iteration of the steps of a MSD protocol generates pure single non-stabilizer states, or magic states, from multiple copies of a mixed resource state using stabilizer operations only. Thus mixed resource states  promote the stabilizer operations to full universality. Magic state distillation was introduced for qubit-based quantum computation, but little has been known concerning MSD in higher dimensional qudit-based computation. Here, we describe a general approach for studying MSD in higher dimensions. We use it to investigate the features of a qutrit MSD protocol based on the 5-qutrit stabilizer code. We show that this protocol distills non-stabilizer magic states, and identify two types of states, that are attractors of this iteration map. Finally, we show how these states may be converted, via stabilizer circuits alone, into a state suitable for state injected implementation of a non-Clifford phase gate, enabling non-Clifford unitary computation.
\end{abstract}
\maketitle

\section{Introduction}
Quantum computers hold the promise of solving certain computational tasks at an exponentially faster rate than is currently believed to be possible with classical computers \cite{S99}. One of the main obstacles making the task of building a quantum computer difficult is due to quantum decoherence, where errors can, if not corrected, accumulate and spread rapidly.
The theory of quantum fault-tolerance (FT) \cite{S96ft} provides a means to protect the  coherent quantum state and allow a reliable quantum computation. This can be achieved provided the physical error rate is below a certain threshold value. The exact value of the threshold error depends on the fault-tolerance scheme adopted for the computation. 

All FT schemes have a limited set of operations allowing direct implementation on the encoded quantum information \cite{EK09}, and in most known schemes these are the stabilizer operations. Stabilizer operations consist of a family of unitary circuits known as the Clifford group, preparation of $\ket{0}$ state and measurement in the computation basis. Although these operations can produce highly entangled states the Gottesman-Knill theorem \cite{NC00} tells us that a computation consisting of the stabilizer operations alone can be efficiently simulated by a classical computer. The stabilizer state operations are not quantum universal; in fact, Clifford circuit operations alone cannot even implement non-linear classical logic gates, such as the Toffoli gate. Nevertheless, in most quantum FT schemes the stabilizer operations are those which can be most readily achieved fault tolerantly. It is natural then to ask, what additional resources are needed to promote the stabilizer operations to universality?

One answer to this question is given by the theory of magic state distillation (MSD) \cite{K04,BK05msd}. Almost any single qubit gate is approximately universal for SU(2) \cite{lloyd}, and similarly, if the stabilizer operations are augmented by a supply of many copies of almost any ancillary pure state, the states can be used as a resource for the implementation of non-Clifford gates via ``state-injection'' methods \cite{BK05msd}.   A supply of qubits in a particular state may suffice, provided it is not a stabilizer state.

\begin{figure} 
    \includegraphics[scale=.4]{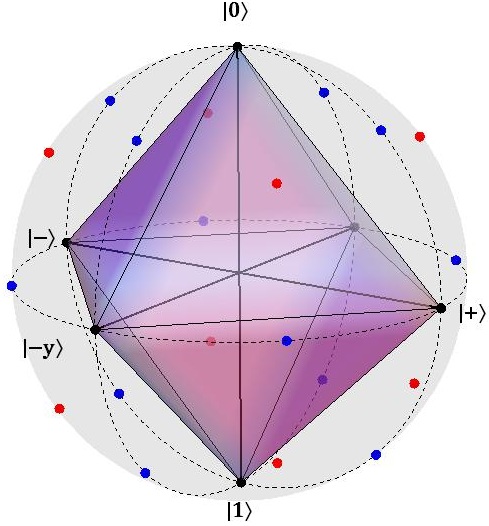}
    \caption{The Bloch sphere with the qubit stabilizer octahedron and the magic states. There are $12$ (blue) \textit{H}-type magic states and $8$ (red) \textit{T}-type magic states.}
    \label{magicstates}
\end{figure}

To compute fault tolerantly, we require a scheme that tolerates preparation noise for these ancillary resources.  Hence, we are interested in which mixed quantum states provide a suitable resource. One can straight away identify a set of states which will \emph{not} be useful in this regard. These are the single qubit states that can be prepared via the stabilizer operations, together with classical randomness. These states are the convex hull of the Pauli eigenstates, and we shall call such states the \textit{stabilizer states}. In the Bloch picture they occupy an octahedron, as shown in Fig. (\ref{magicstates}). 

\earl{Despite these obstacles, Bravyi and Kitaev~\cite{K04} showed that some mixed non-stabilizer states can enable universal quantum computing.  They proposed a process of ``magic state distillation" whereby suitable resource states are efficiently converted, using only stabilizer operations, into a smaller number of purer non-stabilizer states, the so-called magic states.  While their methods are extremely powerful, they only considered the qubit case. For non-generic noise models, it is also possible to \hus{distil} certain 3-qubit states that can be used to implement the Toffoli gate \cite{D01}.  Here we tackle the problem of distilling pure 3-dimensional, or qutrit, non-stabilizer states, and herein refer to all pure non-stabilizer states as magic states.}

\earl{Known magic state distillation schemes have an iterative structure \cite{BK05msd,CB09}, with each iterate having 3 steps: 
\begin{enumerate}
 \item \textit{Initialization},  prepare $n-$copies of the qudit resource state $ \rho_{r} $;
 \item \textit{Projection},  measure Pauli-observables that stabilize a $d$-dimensional subspace and postselect on the $+1$ outcome;
 \item \textit{Decoding}, perform a Clifford unitary that maps the $d$-dimensional subspace onto a single physical qudit $\rho^{out}_{r}$.
 \end{enumerate}
When successful, the output state $\rho^{out}_{r}$ is used as one of the inputs on the next level of iteration.  All known magic state distillation protocols achieve higher purities via iteration, although non-iterative protocols are an interesting possibility;  for example, compare with the hashing protocol~\cite{hashing1,DevWinter} and quantum polar codes \cite{RDR11} used in analogous context of entanglement distillation.  If there exists a protocol that iteratively, or by other means, converts a resource state $ \rho_{r} $ into a magic state of arbitrarily high purity, then $ \rho_{r} $ is said to be resource state.}

Magic state distillation for qubits has an elegant geometrical visualisation in terms of the Bloch sphere.  \earl{In this representation}, the stabilizer states form an octahedron, whose vertices are the Pauli eigenstates. Clifford Group unitaries coincide with symmetries of this octahedron. The magic states distilled by the protocol of \cite{BK05msd} correspond to the states invariant under certain rotational symmetries. The first of these is the set of $180^{\circ}$ rotations around the edges of the octahedron, which contains the Hadamard gate, and we shall call Hadamard-type, or H-type rotations. The second type are $120^{\circ}$ rotations around the faces of the octahedron, \earl{known as T-type rotations.}
\begin{figure} 
    \includegraphics[scale=.8]{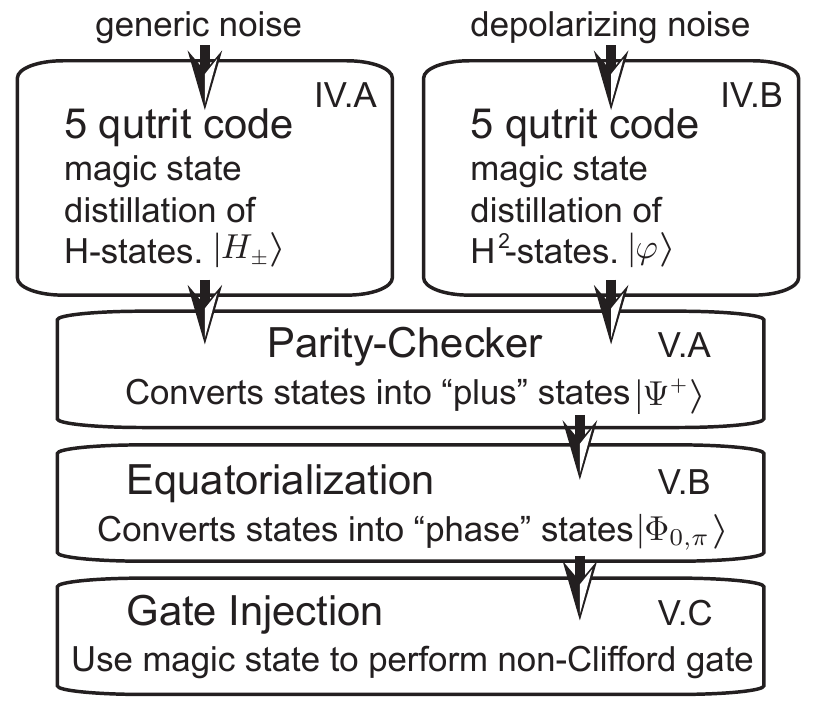}
    \captionsetup{singlelinecheck=yes, justification=centerlast}
      \caption{An overview  of how different protocols in this paper are related.  We consider distillation of two different classes of states using the 5-qutrit code.  The output of either can be converted by 2 subsequent sub-protocols, called parity-checking and equatorialization, to produce different magic states.  The output of equatorialization can then be exploited to perform a non-Clifford unitary.  The relevant section is noted in the top-right corner of each protocol box.}
    \label{fig_overview}
\end{figure}

There has been a considerable amount of  work to improve the original qubit schemes in \cite{BK05msd} and to modify their noise model. Reichardt \cite{R05css} showed how all the states above the edges of the octahedron can be distilled by Steane's 7 qubit code. \earl{In contrast, two interesting no-go theorems for qubit and odd-dimensional systems show that not all non-stabilizer states are useful resources.  Recently, Veitch \textit{et. al.} \cite{V12} showed that for odd-dimensional systems there exist bound states that cannot contribute any enhancement to the computational power  of the stabilizer operations.  The qubit problem is more subtle. Campbell and Browne \cite{CB10bs} showed that for any iterative protocol there will always exist undistillable qubit states above the faces of the octahedron.  However, non-iterative qubit protocols are still poorly understood.}  Furthermore, Campbell \cite{C10ca} introduced an activation protocol that can activate \earl{qubit states from above the octahedron face} to the distillable regions. The magic state distillation protocol allows us to upper bound the error tolerance threshold \cite{PV10} and is at the heart of state-of-the-art fault tolerance schemes  \cite{RHG06}. Moreover, an experimental implementation of MSD has been recently demonstrated in an NMR system for the $ 5- $qubit code \cite{SZRL}.

Surprisingly, very little study has been made of whether magic state distillation can be generalized to higher dimensions. There are a number of reasons why such a study would be important.   In particular, topological quantum systems have anyons with braiding statistics and a dimensionality that is fixed by the physical system.  Research into understanding these systems is ongoing \cite{G06,AGF09}, but it is entirely plausible that the most promising systems provide braiding statistics corresponding to Clifford operations of dimensions greater than 2. Furthermore, the original magic state schemes have some surprising features (hence the name ``magic states''), and it is unclear how much these depend on special features of the qubit Clifford group, or whether they are generic. There are therefore many open questions.  Can the region of states which converge under distillation be identified? A study of higher dimensional MSD may give new insights into the structural differences between the Clifford group in two-dimensions, which plays a central role in quantum computation theory, and in higher-dimensions, where it has been intensively studied in the context of SIC-POVMS \cite{A05}, MUBs \cite{ABC08,A09} and DWF (discrete Wigner functions) \cite{G06}.

Recently, expanding on their work in \cite{DH09nt}, van Dam and Howard \cite{DH11dwf} have studied noise thresholds with $d$-dimensional quantum systems. They have calculated the set of robust qudit states that are most resilient to depolarizing noise and found the degree of noise needed to map such states to the set of  stabilizer states increases with dimension scaling with $d/(d+1)$. Thus, the higher dimensional states have the potential to offer higher magic state distillation thresholds. Until now, however, no magic state distillation schemes for $d>2$ had been presented.

In this paper, we present the first generalization of magic state distillation to higher dimensional systems. We demonstrate that magic state distillation can be achieved in a qutrit system, and find both similarities and differences with previously known MSD protocols for qubits. We have chosen to focus on the three-dimensional qutrit space for this first study since it has the benefit of prime dimension, and computational tractability.  Nevertheless, many of the features which we uncover are likely to be generic.

\hus{We start in Sec. \ref{secDef} by defining notation and discussing the state space and the Clifford group in higher dimensions. In Sec. \ref{secmsd5} we present a generalised approach to study the distillation properties of any stabilizer code of any prime dimension.}  The main protocol we consider is a qutrit generalization of the $ 5$-qubit code covered in Sec. \hus{\ref{5qutritcode}} and App. \ref{appdistillation}. Here we find magic state distillation occurring for 2 distinct families of states. The first family contains a pair of eigenstates of the qutrit Hadamard operator. Under depolarizing noise, they are distilled up to an error threshold of $23.3\%$. This family is generic in that all quantum states can be mapped into the Hadamard plane by random application of the Hadamard unitaries, a process known as Hadamard-twirling. The second family contain four eigenstates of the qutrit Hadamard-squared operator, but lies within a degenerate eigenspace of this operator, and so is not uniquely defined by it. Under depolarizing noise, they are distilled up to an error threshold of $34.5\%$.

Not all magic states, such as those output by the 5-qutrit code, are known to be useful for state injected non-Clifford gates. However\hus{,} certain special magic states, which we call phase-states, do have the capacity to implement non-Clifford unitaries.  We show that, by executing additional protocols, the outputs from the 5-qutrit code can be converted into the desirable phase-states.  We overview the whole process in Fig.~(\ref{fig_overview}), which shows that the promotion of the Clifford group can be achieved by following 5-qutrit distillation by parity-checking and equatorialization, which we describe in Sec.~\ref{non-Clifford}.

\section{Definitions and Notation}\label{secDef}

A quantum dit, or a qudit, is a $ d- $level quantum system where $ d \geq 2 $. The corresponding quantum state $ \rho $ can be represented by a positive semidefinite $d\times d$ matrix of unit trace. In the language of linear algebra, such a matrix can be decomposed as a linear sum of a basis set. For our purposes we have adopted the Weyl basis (also known as the Heisenberg-Weyl basis). This is because the Weyl basis set is a natural generalization of the conventional qubit Pauli operators which makes them very convenient in studying MSD. Herein we assume that $ d $ is a prime dimension.
\subsection{Higher Dimensions and Stabilizer Codes}
The $ d- $dimensional single-qudit $ X $ and $ Z $ Pauli operators are defined as \cite{G99}:
\begin{equation}\label{genxz}
X=\displaystyle\sum^{d-1}_{j=0}\ket{(j+1)\hspace{-2mm}\mod d}\bra{j}, \hspace{2mm}
Z=\displaystyle\sum^{d-1}_{j=0}\omega^{j}\ket{j}\bra{j}, 
\end{equation}
where $ \omega=e^{2\pi i/d} $ is the $ d$-th root of unity. From this definition we see that $ X $ and $ Z $ are traceless non-Hermitian unitaries. They obey the commutation relation $ XZ=\omega^{-1}ZX $ and are cyclic in the power of $ d $ (i.e $X^{d}=Z^{d}=I$). We define a slightly different form of the single-qudit Pauli operators as follows:
\begin{equation}\label{defbases}
\{\sigma_{j,k}=\omega^{cjk}X^{j}Z^{k} ;\hspace{1mm} (j,k)\in\mathbb{Z}_{d}^{2}\},
\end{equation}
where $  c=(1-d)/2$. The main reason for choosing this definition and the significance of the extra phase $ \omega^{cjk} $ will become apparent in section \ref{secquditspace}. But we shall first outline some of the properties of the Pauli operators based on this definition. For a single qudit, the composition of two Pauli operators can easily be verified to be
\begin{equation}\label{composition}
\sigma_{j,k}\sigma_{j',k'}= \omega^{j'k-c(jk'+j'k)}\sigma_{(j+j'),(k+k')}.
\end{equation}
For the case of a composite system of $n-$qudits we use the symplectic notation to represents the  $n-$fold tensor products of Pauli operators
\begin{align}\label{nfold}
\sigma_{j_{1},k_{1}}\otimes\sigma_{j_{2},k_{2}}\otimes\dots\otimes\sigma_{j_{n},k_{n}}&\equiv\sigma_{j_{1}j_{2}\dots j_{n},k_{1}k_{2}\dots k_{n}}\notag\\
&\equiv\sigma_{\boldsymbol{j},\boldsymbol{k}},
\end{align}
where $\boldsymbol{j}$ and $\boldsymbol{k}$ are vectors in $ \mathbb{Z}_{d}^{n} $. The  Pauli operators satisfy a generalized commutation relation
\begin{equation}\label{comrel}
\sigma_{\boldsymbol{j},\boldsymbol{k}}\sigma_{\boldsymbol{j}',\boldsymbol{k}'}=\omega^{\boldsymbol{k}.\boldsymbol{j}'-\boldsymbol{j}.\boldsymbol{k}'}\sigma_{\boldsymbol{j}',\boldsymbol{k}'}\sigma_{\boldsymbol{j},\boldsymbol{k}},
\end{equation}
where $ \boldsymbol{k}.\boldsymbol{j}'-\boldsymbol{j}.\boldsymbol{k}' $ is the symplectic inner product. Based on the above definitions, the $ n- $qudit Pauli group $ \mathcal{P}_n $ is
\begin{equation}\label{quditstabgroup}
\mathcal{P}_{n}=\{\omega^{l}\sigma_{\boldsymbol{j},\boldsymbol{k}}\hspace*{1mm}|\hspace*{1mm} l\in\mathbb{Z}_{d}\}
\end{equation}
All element of $ \mathcal{P}_{n} $ have eigenvalues of the form $ \omega^{m} $ for $ m\in\mathbb{Z}_d$.

We will now briefly \hus{review} stabilizer formalism and the basic properties of stabilizer codes \cite{G97phd}.  A stabilizer state $ \ket{\psi} $ is a simultaneous $ +1 $ eigenvector of an Abelian Pauli subgroup $ \mathcal{S}_{n}\in\mathcal{P}_n $ called the stabilizer group. The subgroup $ \mathcal{S}_{n} $ is generated by $ n $ independent and mutually commuting generators $ \left< g_{1},g_{2},\dots,g_{n-k}\right>$ with $ |\mathcal{S}_{n}|=d^{n} $ and \hus{$ \omega^{m}I\notin\mathcal{S} _{n}$} for all non-zero $ m\in\mathbb{Z}_{d} $. 
The stabilizer group $ \mathcal{S}_{n} $ forms a code, called the stabilizer code, with the stabilizer states being the codewords in the code-space. Stabilizer codes encodes $ k $ logical qudits ($ d^{k}-$dimensional Hilbert space $ \mathcal{H}_{k} $) into a larger Hilbert space $ \mathcal{H}_{n} $ of $ n $ physical qudits. We denote such code by $ [[n,k,\delta]]_d $, where $ \delta $ is the distance of the code and the subscript $ d $ is the dimension of the Hilbert space.
Finally, we represent the logical operators on the code subspace by $  \sigma_{j,k}^{L} $, such that $\sigma_{j,k}^{L} \in\mathcal{P}_{n}\backslash \mathcal{S}_{n}$ and commutes with all elements of $ \mathcal{S}_n $.

\subsection{Qudit Space and Pauli Group Orbits}\label{secquditspace}
When we discuss the general structure of a magic state distillation in Sec. \ref{secmsd} we will see that it is most convenient to use $ \sigma_{j,k} $ as the basis set to represent a qudit state $ \rho $. But recall that $ \sigma_{j,k} $ is a \textit{non-Hermitian} unitary operator.  To guarantee the Hermiticity of $ \rho $ we must therefore impose the Hermiticity  condition $ \rho=\rho^{\dagger} $. We begin by expressing $ \rho $ as
\begin{equation}\label{rhosigma}
\rho(\boldsymbol\alpha)=\frac{1}{d}\displaystyle\sum_{(j,k)}\alpha_{j,k}\sigma_{j,k}.
\end{equation}
where the summation is over all pair elements of $ \mathbb{Z}_{d}^{2} $ and we have assumed that $ \alpha_{0,0}=1 $. We will use $ \sigma_{0,0} $ and the conventional $ \openone $ interchangeably. We refer to $ \alpha_{j,k} $ as the Bloch components, as a generalization of the qubit convention, and $ \boldsymbol{\alpha} $ as the Bloch vector which has the Bloch components as its elements. Observe that for $d=2$, the set of  $ \sigma_{j,k} $ is Hermitian and the Bloch components will be real, but in the general qudit case the Bloch components are complex, constrained by a Hermiticity relation in order for $ \rho=\rho^{\dagger} $ to hold. To work out the Bloch components' relation we start by explicitly writing $ \rho=\rho^{\dagger} $: 
\begin{equation}\label{rhodagger}
\displaystyle\sum_{(j,k)}\alpha_{j,k}\sigma_{j,k} = \displaystyle\sum_{(j,k)}\alpha_{j,k}^{*}\sigma_{j,k}^{\dagger}.
\end{equation}
The importance of the extra phase factor of $ \omega^{cjk} $ in our definition in \hus{Eq.} (\ref{defbases}) is to ensure that $ \sigma_{j,k}^{\dagger}=\sigma_{-j,-k} $, which can be easily verified. Using this fact, and after relabelling, \hus{Eq.} (\ref{rhodagger}) reduces to
\begin{equation}\label{hermcon}
\alpha_{j,k}^{*}=\alpha_{-j,-k}.
\end{equation}
A direct implication of \hus{Eq.} (\ref{hermcon}) is that only half the Bloch components, or $(d^{2}-1)/2 $, are \textit{independent}, as the other half are simply the complex conjugates. Hence, only half the Bloch components are needed to define the density operator. Of course, the independent Bloch components are complex and we still have $(d^{2}-1) $ real parameters defining the density operator. A Bloch component $ \alpha_{j,k} $ can be evaluated using the following relation:
\begin{equation}\label{blochcomp}
\alpha_{j,k}=\mbox{tr}(\rho\sigma_{j,k}^{\dagger})=\mbox{tr}(\rho\sigma_{-j,-k}).
\end{equation}

Beyond the qubit case, it is not possible to visualise the entire state space with a geometrical picture similar to the Bloch sphere. However, there have been some attempts to study the geometry of the state space in higher dimensions \cite{BD08}, and for a Bloch type representation for qutrits \cite{K09,GS11}. \earl{Having fixed normalization and Hermiticity, a mixed qutrit state is described by a complex vector $\vec{\alpha} \in \mathbb{C}^{(d^{2}-1)/2}$.  This complex vector space has a natural inner product, $\langle \vec{\alpha}, \vec{\beta} \rangle=  \sum_{j}\alpha_{j}^{*}\beta_j{}$, a norm $| \vec{\alpha} | = \sqrt{ \langle \vec{\alpha}, \vec{\alpha} \rangle }$, and distance $D(\vec{\alpha} , \vec{\beta}) = | \vec{\alpha} - \vec{\beta}|$.   These geometric concepts are related to the density matrix representation via
\begin{eqnarray}
	\tr ( \rho_{\vec{\alpha}}^{\dagger} \rho_{\vec{\beta}} )  	& = & ( 1+2 \langle \vec{\alpha} , \vec{\beta} \rangle ) / d.
\end{eqnarray}
Since all pure state satisfy $\tr ( \rho_{\vec{\alpha}} ^{2} ) \leq 1 $, this entails
\begin{equation}
	| \vec{\alpha }|^{2} \leq \frac{d-1}{2},
\end{equation}
and all physical states are within a Bloch-like ball of radius $\sqrt{(d-1)/2}$ about the origin, with pure states on the surface of the Bloch ball.}  The qubit state space is of course a special case in which all the points on the surface of the Bloch sphere corresponds to positive physical states. The additional condition required to ensure a positive pure state, as shown in \cite{JL05}, is  $ \mbox{tr}(\rho^{2})=\mbox{tr}(\rho^{3})=1 $.

\earl{The geometry vectors $\vec{\alpha}$ corresponding to physical states $\rho_{\vec{\alpha}}$ is quite intricate.  However, within certain hyperplanes the substructure is very simple.  Consider hyperplanes defined by a set of $d$ positive orthonormal operators, $\{ \rho_{\vec{\beta}_{j}} \}$, such that $\tr ( \rho_{\vec{\beta}_{j}}  \rho_{\vec{\beta}_{k}}) = \delta_{j,k}$.   Note that in the geometric picture, this entails
\begin{eqnarray}
	\langle \vec{\beta}_{j} , \vec{\beta}_{k} \rangle & = &  \frac{1}{2}( d \delta_{j,k}-1) .
\end{eqnarray}
We consider the complex hyperplane spanned by $\{ \vec{\beta}_{j} \}$, such that $\vec{\alpha} = \sum_{j} b_{j} \vec{\beta}_{j}$.  It follows that an operator $\rho_{\vec{\alpha}}$ is positive if and only if $\sum_{j} b_{j} \leq 1$ and $b_{j}=|b_{j}|$ for all $j$.  Hence, the physical $\vec{\alpha}$ lie within the convex polytope with $\vec{\beta}_{j}$ as vertices.  We have $d$ vertices all equally separated from each other and residing within a real $d-1$ dimensional hyperplane.  Hence, the polytope has the structure of a standard simplex.  For $d=3$, and an appropriate plane, the physical states reside within an equilateral triangle.}


Finally, we shall discuss the orbits of the Pauli group $ \mathcal{P}_{n} $ when acting on a general qudit state $ \rho(\boldsymbol{\alpha}) $ with conjugation being the group action. The singular orbit of a general state $ \rho(\boldsymbol{\alpha}) $, denoted by $ \mbox{Orb}(\rho(\boldsymbol{\alpha})) $, is defined as
\begin{equation}
\mbox{Orb}(\rho(\boldsymbol{\alpha}))=\{\rho(\boldsymbol{\alpha '})=\sigma_{\boldsymbol{j '},\boldsymbol{k '}} \hspace{1mm}\rho(\boldsymbol{\alpha})\hspace{1mm}\sigma_{\boldsymbol{j '},\boldsymbol{k '}}^{\dagger}\hspace{2mm}\forall\hspace{.5mm} \sigma_{\boldsymbol{j '},\boldsymbol{k '}}\in \mathcal{P}_{n}\}.
\end{equation}
In our Bloch representation we are using Pauli group elements as a basis set for the states, thus conjugation by Pauli operators will not transform the basis elements, but will add a phase of the form $ \omega^{l} $ for some $ l\in\mathbb{Z}_d $. Therefore, the overall effect of this conjugation is to add certain phases to the Bloch components. The exact form of the phases is exactly given by:
\begin{align}
\rho(\boldsymbol{\alpha'})&=\displaystyle\sum_{(\boldsymbol{j},\boldsymbol{k})}\alpha_{\boldsymbol{j},\boldsymbol{k}}\sigma_{\boldsymbol{j'},\boldsymbol{k'}}\sigma_{\boldsymbol{j},\boldsymbol{k}}\sigma_{\boldsymbol{-j'},\boldsymbol{-k'}},\\
&=\displaystyle\sum_{(\boldsymbol{j},\boldsymbol{k})}\omega^{\boldsymbol{j}\boldsymbol{k'}-\boldsymbol{j'}\boldsymbol{k}}\alpha_{\boldsymbol{j},\boldsymbol{k}}\sigma_{\boldsymbol{j},\boldsymbol{k}}\label{orb}.
\end{align}
where $ \sigma_{\boldsymbol{j'},\boldsymbol{k'}}^{\dagger} =\sigma_{\boldsymbol{-j'},\boldsymbol{-k'}}$, the commutation and composition relations where used in the last step.
\subsection{Qutrits}
So far the discussion has been for all prime dimensions, but in the remainder of this paper we will discuss the qutrit case only. Therefore, we shall outline some of above results explicitly for the $ d=3 $ case. Our definition of the qutrit Pauli basis set in \hus{Eq.} (\ref{defbases}) is $\sigma_{j,k}=\omega^{-jk}X^{j}Z^{k}$, where $ \omega=e^{2\pi i/3} $ and $ c=-1 $. The explicit qutrit $ \rho(\boldsymbol\alpha) $ state is:
\begin{align}
&\rho(\boldsymbol\alpha)=\frac{1}{3}\Bigl(\sigma_{0,0}+\alpha_{1,0}\sigma_{1,0}+\alpha_{1,0}^*\sigma_{2,0}+\alpha_{0,1}\sigma_{0,1}\notag\\ 
&+\alpha_{0,1}^*\sigma_{0,2}+\alpha_{1,1}\sigma_{1,1}+\alpha_{1,1}^*\sigma_{2,2}+\alpha_{1,2}\sigma_{1,2}+\alpha_{1,2}^*\sigma_{2,1}\Bigr).
\end{align}
As we can see, completely specifying a qutrit state would only require $ 4 $ complex independent parameters.  In terms of the Bloch components, the purity condition $ \mbox{tr}(\rho^{2})=1 $ for a general qutrit state can be shown to be $|\vec{\alpha}| \leq 1$.


The Pauli group orbits for a single qutrit state can be evaluated using \hus{Eq.} (\ref{orb}). We are interested in knowing how the phases of the four independent Bloch components change when an element from the $ 9 $ qutrit $ \sigma_{j,k} $ operators is conjugated with the general qutrit state. The result is summarised in table \ref{phasestabel}. We refer to these phases as the orbital Bloch phases. These represent the phases which generate the set of states Pauli equivalent to any state. The magic states that we will find are unique up to a Bloch orbital phase. In other words, inserting one of the phases from the set in table \ref{phasestabel} into the Bloch components of the magic states would also give a valid magic state with the same distillation properties.
\begin{center}
\begin{table}
\begin{tabular}{c|c}
$ \sigma_{\boldsymbol{j'},\boldsymbol{k'}}  $ &$\sigma_{\boldsymbol{j'},\boldsymbol{k'}}\rho(\alpha_{1,0},\alpha_{0,1},\alpha_{1,1},\alpha_{1,2})\sigma_{\boldsymbol{j'},\boldsymbol{k'}}^{\dagger} $ \\ \hline
$ \sigma_{ 0 ,0}  $&  $ \rho(\alpha_{1,0},\alpha_{0,1},\alpha_{1,1},\alpha_{1,2}) $\\
$ \sigma_{ \pm 1,0}  $& $ \rho(\alpha_{1,0},\omega^{\mp 1}\alpha_{0,1},\omega^{\mp 1}\alpha_{1,1},\omega^{\pm 1}\alpha_{1,2}) $ \\
$ \sigma_{0,\pm 1}  $ & $ \rho(\omega^{\pm 1}\alpha_{1,0},\alpha_{0,1},\omega^{\pm 1}\alpha_{1,1},\omega^{\pm 1}\alpha_{1,2}) $ \\
$ \sigma_{\pm 1,\pm 1} $ & $ \rho(\omega^{\pm 1}\alpha_{1,0},\omega^{\mp 1}\alpha_{0,1},\alpha_{1,1},\omega^{\mp 1}\alpha_{1,2}) $ \\
$ \sigma_{\pm 1,\mp 1} $ & $ \rho(\omega^{\mp 1}\alpha_{1,0},\omega^{\mp 1}\alpha_{0,1},\alpha_{1,1}\omega^{\pm 1},\alpha_{1,2}) $ \\
\end{tabular}
\caption{The qutrit orbital Bloch phases.}
\label{phasestabel}
\end{table}
\end{center}

\subsection{Stabilizer Dynamics and the Clifford Group}

One the most important advantages of the stabilizer formalism is the simplicity it provides when studying the dynamics of the stabilizer states. Instead of studying the action of a unitary $ U $ on a $ n- $qudit stabilizer state $ \ket{\psi_{s}} $ (which would require the complete description of the map on $ d^{n} $ parameters), the problem can be reduced to studying the evolution of the stabilizer operators by $ U $ in the Heisenberg picture (which is linear in $ n $) \cite{G98}. An important class of unitaries are those that map the elements of the stabilizer group back to the stabilizer group under conjugation. These operation are called the Clifford unitaries. More formally, the Clifford group is the normalizer of the stabilizer group, defined as
\begin{equation}
\mathcal{C}_{n}=\{C | CP_{i}C^{\dagger}=P_j \hspace*{2mm}\forall \hspace*{1mm}P_{i},P_{j}\in \mathcal{P}_{n}\}.
\end{equation}

For any prime dimension, the Clifford group has been shown to be generated by three gates \cite{G99,C06}. These are the Hadamard gate $ H $, the Phase gate $ S $ and the Controlled-NOT gate $ \Lambda(X) $. \hus{The $ d- $dimensional Hadamard gate is given by:}
\begin{equation}\label{hgate}
H\ket{j}=\frac{1}{\sqrt{d}}\displaystyle\sum_{k\in \mathbb{Z_{d}}}\omega^{jk}\ket{k}.
\end{equation}
Under conjugation, the Hadamard gate transforms Pauli operators as
\begin{equation}
H\sigma_{j,k}H^{\dagger}=\sigma_{-k,j}.
\end{equation}
For completeness\hus{,} the $ d- $dimensional $ S $ and $ \Lambda(X) $ gates are given by:
\begin{align}
S\ket{j}=&\hspace*{1mm} \omega^{\frac{j}{2}\left(j-1\right)}\ket{j},\\
\Lambda\left(X\right)\ket{j}\ket{k}=&\ket{j}\ket{\left(j+k\right)\hspace*{-2mm}\mod d}.
\end{align}
In addition to unitary dynamics we allow for so-called Pauli measurements.  For qubit systems the Pauli group contains Hermitian operators where for qudit they are unitary but not Hermitian, with the exception of the identity.  Rather, when we say we measure a non-trivial Pauli $P$ this is taken to mean that we perform a POVM measurement that has the eigenvectors of $P$ as POVM elements.

\section{Magic State Distillation} \label{secmsd5}

Using the definitions and notations we have developed in the previous section, we will show how the three steps of a MSD protocol described in the introduction can be formulated to study the distillation properties of any stabilizer code of any prime dimension. 

\textbf{Resource state preparation:} The computational model considered when studying MSD consists of perfect stabilizer operations and the ability to prepare $ n $ identical copies of a noisy resource state $ \rho_{r}$. It should be noted that the resource state is noisy due to the imperfect preparation process. By repeating the preparation procedure $ n $ times, the state $ \rho^{\otimes n}_{r}$ will be prepared. As an input to the MSD protocol we consider a general state 
\begin{equation}\label{nrho}
\rho^{\otimes n}=\frac{1}{d^n}\displaystyle\sum_{(\boldsymbol j,\boldsymbol k)\in\mathbb{Z}_{d}^{n}}\alpha_{j_{1}\dots j_{n},k_{1}\dots k_{n}}\sigma_{j_{1}\dots j_{n},k_{1}\dots k_{n}}.
\end{equation}
By performing the remaining steps of the iteration on the above general form, we will determine the map on the Bloch components of the initial general state $ \rho $. Then, by searching the state space for different initial states, we can identify the resource states as those that when used as an input to the protocol the output state has a higher fidelity with respect to a pure non-stabilizer state, and ultimately distilling this non-stabilizer pure state. If the search is done systematically, one can in principle identify the entire region of resource states $ \rho_{r} $. 


\textbf{Stabilizers measurement and Decoding:}  The $ (n-k) $ stabilizer generators of a stabilizer code $ [[n,k,\delta]]_{d} $ are measured successively along with postselecting on the $ +1 $ outcome of each measurement. That is, if one of the outcomes is $ \omega^{k} $ (for some non-zero $ k\in\mathbb{Z}_{d} $) then the protocol is aborted, and the procedure is repeated with a fresh state $ \rho^{\otimes n} $. Also, it is important to notice that the error correction code is not being used for the usual purpose of correcting errors since the syndrome measurements are performed on the product state $ \rho^{\otimes n} $. If successful, the measurement of the stabilizers simply project the state to the code's subspace. The projector operator describing this measurement procedure can be put into a convenient form to us as
\begin{equation}\label{genproj}
\Pi=d^{k-n}\displaystyle\sum_{\boldsymbol m \in \mathbb{Z}_{d}^{n-k}}g_{1}^{m_{1}}g_{2}^{m_{2}}\dots g_{n-k}^{m_{n-k}}.
\end{equation}
After the measurements, the $ n $ copies of $ \rho $ will be projected into the code's subspace, and the following map will be performed:
\begin{equation}
\rho^{\otimes n}\mapsto\frac{\Pi\rho^{\otimes n}\Pi^{\dagger}}{\mbox{tr}\left(\rho^{\otimes n}\Pi\right)}.
\end{equation}
The state is decoded via a Clifford operator \cite{CB09}. In a Heisenberg picture, the decoding operation maps logical operators on the code-space to unencoded operators acting on a single qudit. The output Bloch components after decoding $ \alpha_{j,k}^{out} $ therefore corresponding to the components of the logical operators prior to decoding $ \sigma_{j,k}^{L} $. After one round of the distillation, these can be evaluated as follows:
\begin{equation}
\alpha_{j,k}^{out}=\frac{\mbox{tr}(\Pi\rho^{\otimes n}\Pi^{\dagger}(\sigma_{j,k}^{L})^{\dagger})}{\mbox{tr}(\rho^{\otimes n}\Pi)}.\label{genmap1}
\end{equation}
The resultant expressions for the output Bloch components will be multi-variable complex polynomials of order $ n $.  \earl{For the qutrit codes we consider here, we have not found analytic solutions for the fixed points of the map.  However, the problem is tractable by using numerical methods to study the distillation behaviours and the fixed points to a high accuracy.}

\section{The 5-qutrit code}\label{5qutritcode}

Using the generalized formulation of MSD in the previous section we have studied the distillation properties of the five qutrit code. The stabilizer generators of the general five qudit code $ [[5,1,3]]_d $ takes the same form in all dimensions. It is usually presented in terms of the conventional generalized Pauli operators of \hus{Eq.} (\ref{genxz}), as shown in table \ref{XZgen}. 

\begin{table}[h!]
\begin{center}
\begin{tabular}{|lccccc|}
\hline
$ g_{1} =$&$X$&$Z$&$Z^{-1}$&$X^{-1}$&$I$\\
$ g_{2} =$&$I$&$X$&$Z$&$Z^{-1}$&$X^{-1}$\\
$ g_{3} =$&$X^{-1}$&$I$&$X$&$Z$&$Z^{-1}$\\
$ g_{4} =$&$Z^{-1}$&$X^{-1}$&$I$&$X$&$Z$\\
\hline
\end{tabular}
\caption{The stabilizer generators of the five qudit code $ [[5,1,3]]_{d} $ \cite{R96five}.}
\label{XZgen}
\end{center}
\end{table}
Based on these stabilizers we have studied the distillation map of \hus{Eq.} (\ref{genmap}) for the four Bloch components of a general input qutrit state. The exact distillation calculation and the expressions of the distillation map are given in appendix \ref{appdistillation}.  

The decoding of a stabilizer code is not unique, but one of an equivalence class of unitaries, a coset of the  Clifford group, which are all equally valid choices. The choice of decoding will affect the iterative distillation behaviour.  The decoding specified by the logical operators in table~\ref{XZgen} is the canonical one, though we found behaviour was simplified by following each iterate with the following additional Clifford unitary:
\begin{equation}
\label{Rmat}
	R =  \left( \begin{array}{ccc}
	1 & \omega & \omega \\
	\omega^{2} & \omega & \omega^{2} \\
	\omega^{2} & \omega^{2} & \omega \\
	\end{array}  \right) ,
\end{equation}
where this maps a Hermitian operator $\rho_{\vec{\alpha}}$ such that:
\begin{equation}\label{refR}
	\{ \alpha_{1,0} , \alpha_{0,1} , \alpha_{1,1} , \alpha_{1,2} \} _{R}\mapsto  \{ \alpha_{1,2}^{*} , \alpha_{1,1} , \alpha_{0,1}^{*} , \alpha_{1,0} \}.
\end{equation}
Without this corrective Clifford one observes a cycling behaviour throughout the distillation process, see App~\ref{APPcycle}.

We identify two qualitatively different families of states.  Firstly those in the Hadamard plane, which satisfy $H \rho H^{\dagger} = \rho$. Secondly, we investigate distillation of an interesting set of states outside the Hadamard plane.  

\subsection{Hadamard-like Distillation} 

In the qubit case, the eigenstates of the Hadamard gate are known to be magic states, distillable by the five qubit code $ [[5,1,3]]_2 $ \cite{Note1}. Since the exact generalized form of the Hadamard gate is defined in \hus{Eq.} (\ref{hgate}), a good starting point would be to investigate whether the qutrit Hadamard eigenstates can be distilled by $ [[5,1,3]]_3 $. We begin by outlining some of the structural properties of the qutrit Hadamard eigenspace. In the matrix representation the qutrit Hadamard is given by
\begin{equation}
H=\frac{1}{\sqrt{3}}\left(
\begin{array}{ccc}
1 & 1 &1 \\
1 & \omega & \omega^{2}\\
1 & \omega^{2} & \omega\\
\end{array}
\right), 
\end{equation}
where $ \omega^{2\pi i/3} $ and it has the eigenvalues $ (+1,-1,i) $. We label the corresponding three eigenstates as $ (\ket{H_{+}},\ket{H_{-}},\ket{H_{i}})$. The density operators of the eigenstates have the form:
\begin{align}
\ket{H_{+1}}\bra{H_{+1}}&\equiv \rho\left(a,a,b,b\right),\\
\ket{H_{-1}}\bra{H_{-1}}&\equiv \rho\left(b,b,a,a\right),\\
\ket{H_{i}}\bra{H_{i}}&\equiv \rho\left(c,c,c,c\right),
\end{align}
where $a= \frac{1}{4}\left(1+\sqrt{3}\right) $, $b= \frac{1}{4}\left(1-\sqrt{3}\right) $  and $ c=-\frac{1}{2} $, are real parameters.  This basis of pure states all lie on the hyperplane of operators of the $\rho(x,x,y,y)$.   Probabilistic mixtures of these states form an equilateral triangle and as reviewed earlier, all points outside this triangle correspond to non-physical operators.  Furthermore, any qutrit state can be projected onto the Hadamard plane by applying the following twirling operation to each copy of the input state $ \rho $:
\begin{equation}
\rho\mapsto\displaystyle\sum_{j=1}^{4}\frac{1}{4}H^{j}\rho H^{j^{\dagger}}
\end{equation}
This twirling operation maps the general Bloch components as
\begin{align}
\alpha_{1,0} \hspace*{2mm}\mbox{and} \hspace*{2mm}\alpha_{0,1}&\mapsto\frac{\mbox{Re}(\alpha_{1,0}+\alpha_{0,1})}{2},\\
\alpha_{1,1}\hspace*{2mm} \mbox{and}\hspace*{2mm} \alpha_{1,2}&\mapsto\frac{\mbox{Re}(\alpha_{1,1}+\alpha_{1,2})}{2}.
\end{align}
As such, we are interested in studying the distillable regions within the Hadamard plane, i.e. the states corresponding to the points inside the triangle. 

\begin{figure}
    \includegraphics[scale=.4]{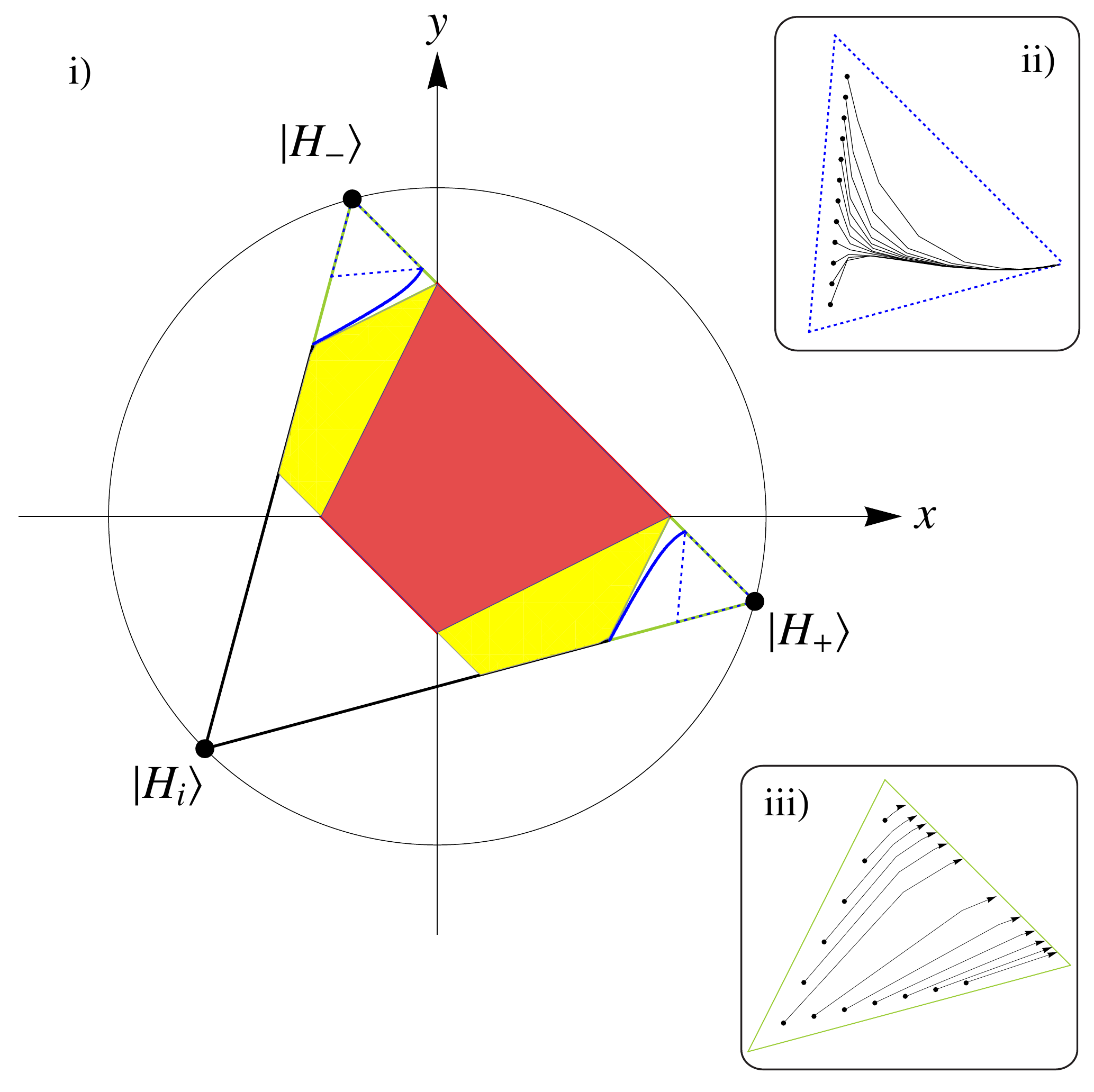}
    \captionsetup{singlelinecheck=yes, justification=centerlast}
      \caption{A representation of the Hadamard plane. The Hadamard eigenstates are the vertices of the equilateral triangle and lies on a circle of radius $ 1/\sqrt{2} $. i) The red region contains the stabilizer states. The yellow and red regions combined form the states with positive Wigner function. The dashed blue and the green triangles contains the states that are distillable by $ [[5,1,3]]_3 $ and $ [[7,1,3]]_3 $ codes, respectively. ii) and iii) shows the distillation paths for the $ \ket{H_{+}} $  state and the mixed states for the $ [[5,1,3]]_3 $ and $ [[7,1,3]]_3 $ codes, respectively.}
      \label{hplane}
\end{figure}

In studying the distillable region in the Hadamard plane, it is also informative to chart out regions for which distillation is impossible by any protocol.  Clearly, all stabilizer states are undistillable (red region in Fig. \hus{(\ref{hplane}i)}), but the results of Veitch \textit{et al} \cite{V12} prove undistillability of all qutrit states with a positive Wigner function. \hus{The numerically calculated positive-region is shown in Fig. \hus{(\ref{hplane}i)} as the yellow area.}

Running the $ [[5,1,3]]_3 $ distillation for all the remaining points as inputs states we have discovered that both the $ \ket{H_{+}} $ and $ \ket{H_{-}} $ states are distillable, but that $\ket{H_{i}}$ is not an attractor. The distillable regions are enclosed by the blue dashed triangles.  The states $ \ket{H_{+}} $ and $\ket{H_{-}}$ are equally valuable as magic states because $R \ket{H_{\pm}} \propto \ket{H_{\mp}}$, where $R$ is a Clifford unitary (see Eq.~\ref{Rmat}), which perhaps also explains the symmetry in their distillation regions. 

The path of the distillation takes the form shown in Fig. \hus{(\ref{hplane}ii)} where we have chosen the $ \ket{H_{+}} $ blue triangle as an example. The small black points are few examples of input states to the distillation, and the black lines represent the distillation paths toward the $ \ket{H_{+}} $ state. Notice how the distillation does not follow a straight line (along the magic axes) as in the qubit case. In fact, in analogy to the qubit case, the above plane is the Hadamard magic plane. The curved distillation path can be understood by studying how the noise of a resource state in the Hadamard plane is suppressed in different directions by the distillation. We start by considering a general state $ \rho_{\medtriangleup} $ inside the triangle of the form
\begin{equation}
\rho_{\medtriangleup}=(1-\epsilon_{1}-\epsilon_{2})\ket{H_{+}}\bra{H_{+}}+\epsilon_{1}\ket{H_{-}}\bra{H_{-}}+\epsilon_{2}\ket{H_{i}}\bra{H_{i}},
\end{equation}
with $ \epsilon_{1}+\epsilon_{2}\le 1 $.  For clarity, we write the Bloch components of the above state as $ \rho_{\medtriangleup}(A_{\medtriangleup},B_{\medtriangleup},C_{\medtriangleup},D_{\medtriangleup}) $. They can be calculated explicitly using \hus{Eq.} (\ref{blochcomp}) as
\begin{align}
A_{\medtriangleup}=B_{\medtriangleup}&=\frac{1}{4}\left(1+\sqrt{3} - 2\sqrt{3}\epsilon_1 - \left(3+ \sqrt{3}\right) \epsilon_{2}\right),\notag\\
C_{\medtriangleup}=D_{\medtriangleup}&=\frac{1}{4}\left(1-\sqrt{3} + 2\sqrt{3}\epsilon_1 - \left(3- \sqrt{3}\right) \epsilon_{2}\right).
\end{align}
Since we know the general distillation map for any set of Bloch components (see App. \ref{appdistillation}), we can simply substitute the above expressions into \hus{Eqs.} (\ref{mapa2}-\ref{mapd2}) to evaluate the output Bloch components $\boldsymbol{\alpha}^{out}=(A_{\medtriangleup}^{out},B_{\medtriangleup}^{out},C_{\medtriangleup}^{out},D_{\medtriangleup}^{out}) $. The output state is then $ \rho^{out}_{\medtriangleup} =\rho(\boldsymbol{\alpha}^{out})$. We have numerically calculated the output $ \epsilon_{1}^{out} $ and $ \epsilon_{2}^{out} $ to the first order term as follows:
\begin{align}
\epsilon_{1}^{out}(\epsilon_1 ,\epsilon_2)&=\bra{H_{-}}\rho_{\medtriangleup}^{out}\ket{H_{-}}\approx\left(0.38+0.09\epsilon_2\right)\epsilon_1,\\
\epsilon_{2}^{out}(\epsilon_1 ,\epsilon_2)&=\bra{H_{i}}\rho_{\medtriangleup}^{out}\ket{H_{i}}\approx\hus{\left(0.77+3.55\epsilon_1\right)\epsilon_2}.
\end{align}
The above expressions shows an asymmetric error suppression in the $ \epsilon_{1} $ (along the $ \ket{H_{+}}-\ket{H_{-}} $ line) and $ \epsilon_{2} $ (along the $ \ket{H{+}}-\ket{H_{i}} $ line) directions. The particular distillation paths of Fig. \hus{(\ref{hplane}ii)} can be explained by observing the difference in the coefficients of $\epsilon^{out}(\epsilon_{1},0) $ and $\epsilon^{out}(0,\epsilon_{2}) $, where we see that in the distillation region of the $ \ket{H_{+}} $ state there is a stronger attraction toward the $ \ket{H_{i}} $ state compared to the $ \ket{H_{-}} $ state.

The above analysis shows that the performance of the $ [[5,1,3]]_3 $ code in distilling the qutrit Hadamard states is not as good as the qubit case where the $ 15 $ qubit code by \cite{BK05msd} has an output error probability of $ \epsilon^{out}\approx 35 \epsilon^{3} $ . This is to be expected given the similar performance of the five qubit code \cite{Note1} in distilling the H-type qubit magic states. 

The state $ \ket{H_{i}} $ is not distillable by $ [[5,1,3]]_3 $. In fact, this state belong to the family of states with maximally non-positive Wigner function~\cite{DH11dwf}. As we can see this state is the furthest away from the stabilizer region in the Hadamard plane and to bring it to the stabilizer region would require a depolarizing noise with an \hus{error threshold of $ 75\% $} (i.e. $ d/(d+1)  $ for $ d=3 $). Whether such a state is distillable is still an open question.  

\begin{table}
\begin{center}
\begin{tabular}{|llllllll|}
\hline
$ g_{1} =$&$I$&$I$&$I$&$X^{-1}$&$X$&$X$&$X^{-1}$\\
$ g_{2} =$&$X$&$I$&$X^{-1}$&$I$&$X^{-1}$&I&$X$\\
$ g_{3} =$&$I$&$X$&$X^{-1}$&$I$&$I$&$X^{-1}$&$X$\\
$ g_{4} =$&$I$&$I$&$I$&$Z$&$Z$&$Z$&$Z$\\
$ g_{5} =$&$Z$&$I$&$Z$&$I$&$Z$&$I$&$Z$\\
$ g_{6} =$&$I$&$Z$&$Z$&$I$&$I$&$Z$&$Z$\\
\hline
$ X_{L} =$&$X$&$X^{-1}$&$X$&$X$&$X^{-1}$&$X$&$X^{-1}$\\
$ Z_{L} =$&$Z$&$Z$&$Z$&$Z$&$Z$&$Z$&$Z$ \\
\hline
\end{tabular}
    \captionsetup{singlelinecheck=yes, justification=centerlast}
\caption{The stabilizer generators of the seven qudit code.}
\label{7gen}
\end{center}
\end{table}

To improve the size of the distillation region we have investigated a qutrit version of the seven qubit code $ [[7,1,3]]_2 $  distillation proposed by \cite{R05css}. We start with the stabilizer generators of $ [[7,1,3]]_2 $ code and by adding  the $ (-1) $ power to the appropriate $ X $ and $ Z $ Pauli operators, we constructed a set of generalized $ 7- $qudit commuting stabilizer generators as shown in table \ref{7gen}. We repeated the distillation procedure for this set of generators for the case $ d=3 $ (exact calculations are omitted here) and we have investigated its distillation capability in the Hadamard plane. We found that this code attracts towards the non-stabilizer segments of the line joining the $ \ket{H_{+}} $ and $ \ket{H_{-}} $ states with the distillation region enclosed by the green triangle in Fig. \hus{(\ref{hplane})}. In other words, the $ 7- $qutrit code distils not pure, but mixed states.   Regardless, the protocol may be useful for bringing states into the region distillable by the 5-\hus{qutrit} code.  The distillation path for the $ \ket{H_{+}} $ state is shown in Fig. \hus{(\ref{hplane}iii)}. This code increase the distillation region as shown by the solid blue curve in Fig. \hus{(\ref{hplane})}. For example, a state between the solid blue line and the dashed blue triangle is first distilled by the seven qutrit code to a state within the dashed blue triangle, after which the $ [[5,1,3]]_3 $ code is used to \hus{distil} the $ \ket{H_{\pm}} $ states.

\subsection{Hadamard-squared subspace} 

\begin{figure}
    \includegraphics[scale=.4]{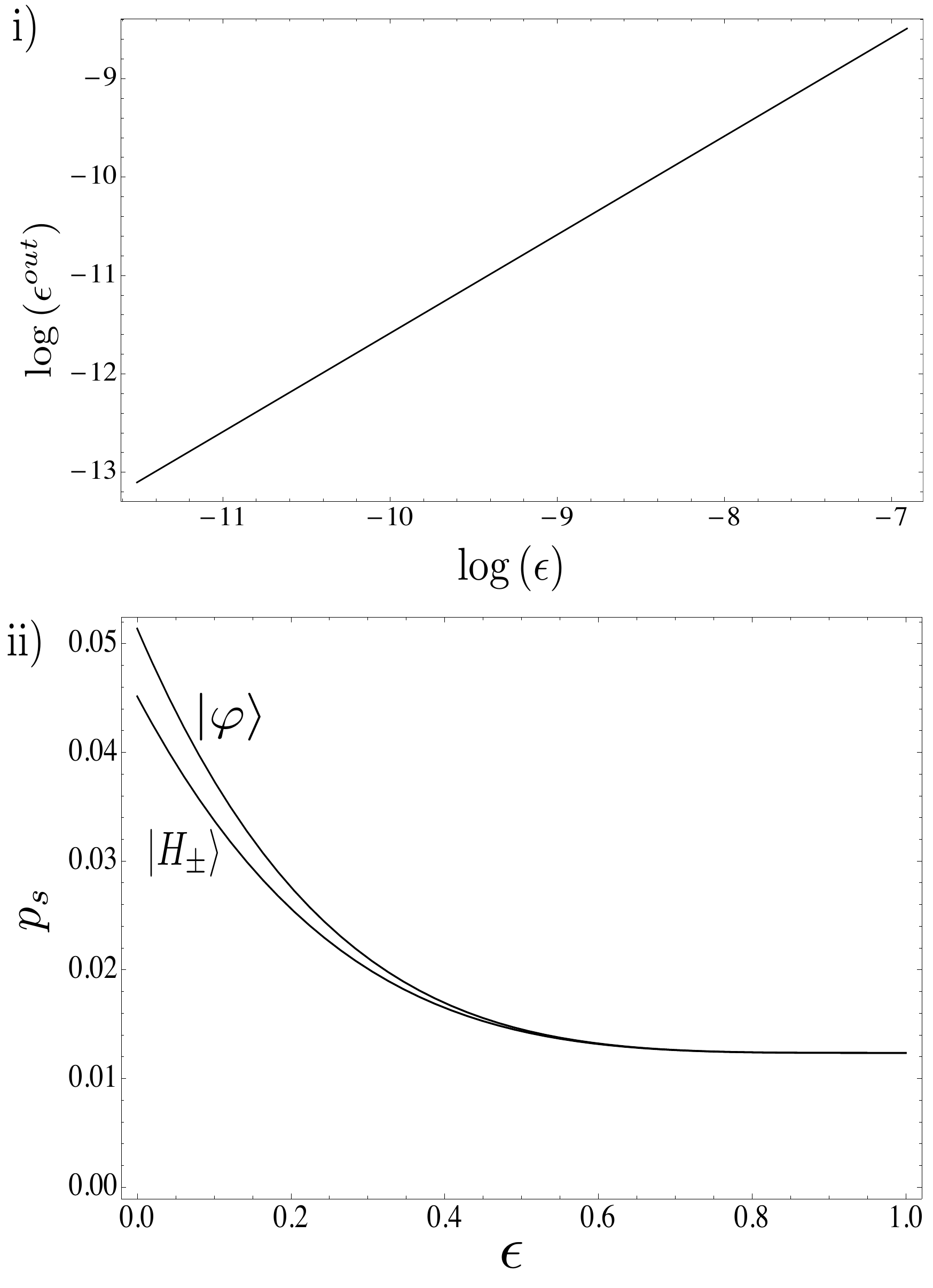}
    \captionsetup{singlelinecheck=yes, justification=centerlast}
      \caption{i) The log-log plot of the output error probability \hus{$\epsilon^{out} $ for the input state $ \rho_{dep} $ and very small depolarizing noise $ \epsilon $}. ii) The success probability for the trivial syndrome measurements for the case where the magic states $ \ket{\varphi} $ and $ \ket{H_{\pm}}$ \hus{are} undergoing depolarizing noise.}
      \label{pf}
\end{figure}

In this section, we introduce a second class of magic states distilled by the $ [[5,1,3]]_3 $ code. The state in question is an eigenstate of the $ H^{2} $ operator, but lies within a degenerate eigenspace for this operator, and so is not uniquely defined by it. The magic state considered here has the form:
\begin{equation}
	\ket{\varphi} = a \ket{0} + b \ket{1} + b \ket{2},
\end{equation}
where up-to 4 decimal places
\begin{eqnarray}
	a & = & -0.1203-0.0272 i , \\ 
	b & = & 0.7017 .
\end{eqnarray}
The equality of the $\ket{1} $ and $\ket{2}$ components follows from the $H^{2}$ symmetry.  In the Bloch representation, the $\vec{\alpha}$ vector is,
\begin{equation}
	\vec{\alpha} = \{  0.3236, -0.4772,  0.5438,  0.6098 \} ,
\end{equation}
with the feature that all components are real being again related to the $H^{2}$ symmetry. Our numerical analysis shows that these states are clearly attractor fixed points of the distillation protocol, \hus{but} we do not have a closed form analytic expressions for them. As a consequence, we will not be able to determine analytically how the error is suppressed as we did in the previous section for the Hadamard magic states. Nevertheless, we can still gain a numerical indication of how the error of $ \ket{\varphi} $ states is suppressed. Lets start with a $ \ket{\varphi} $ state undergoing depolarizing noise:
\begin{equation}
\rho_{dep}=(1-\epsilon)\ket{\varphi}\bra{\varphi}+\epsilon I/3.
\end{equation}
For a sufficiently small $ \epsilon $ the distilled state \hus{$ \rho_{dep}^{out} $} will also be of the above form (i.e on the depolarizing axis). We can then calculate the output error probability as follows:
\begin{equation}
\epsilon^{out}=1-\bra{\varphi}\rho_{dep}^{out}\ket{\varphi}.
\end{equation}
In general, we expect that $ \epsilon^{out}\approx\epsilon^{n} $ for very small $ \epsilon $. Therefore, the power $ n $ can be evaluated as the gradient of a $\log-\log $ plot of $ \epsilon^{out} $ versus $ \epsilon $, as shown in Fig. \hus{(\ref{pf}i)}, and we found that $ n\approx 1 $. 


For completeness, we include the success probability $ p_{\text{s}} $ of the syndrome measurement\hus{s}. Successful syndrome measurement\hus{s}, where all outputs of the stabilizer measurements is $ +1 $, are described by the projector $ \Pi $ given in \hus{Eq.} (\ref{projsimp}). Hence, the probability of this measurement is simply:
\begin{equation}
p_{s}=\text{tr}(\Pi\rho^{\otimes 5 }\Pi)=\text{tr}(\rho^{\otimes 5 }\Pi),
\end{equation}
which is given in \hus{Eq.} (\ref{trrhoproj}) for all sets of Bloch components. We have computed $ p_{s} $ for both $ \ket{\varphi} $ and $ \ket{H_{\pm}} $ undergoing depolarizing noise as an input states to the distillation. A plot of $ p_{s} $ is given in Fig. \hus{(\ref{pf}ii)}. 
\section{Promoting the Clifford group}
\label{non-Clifford}

An interesting practical problem is how to use the magic states, $\ket{H_{+}}$ and $\ket{\varphi}$, or their Clifford equivalent states, to perform injection of a non-Clifford gate.  While their utility is not immediately apparent in their current form, additional sub-protocols can be used to prepare the  \textit{phase}-states that are useful for gate injection. The phase-states are so-called because they only hold phase information, having the following form:
\begin{equation}
	\ket{\Phi_{\theta, \phi}} = \ket{0} + e^{i  \theta} \ket{1} + e^{i  \phi} \ket{2} .
\end{equation}
In Sec.~\ref{secGateInject}, we show how to use these states for gate injection.  However, first we describe, in Sec~\ref{secmsd}, the parity-checking protocol, which is used to convert both $\ket{H_{+}}$ and $\ket{\varphi}$ into the \textit{plus}-state $\ket{\Psi^{+}}=\ket{0}+\ket{1}$.  These plus-states are then input to the equatorialization procedure, in Sec.~\ref{secEquator}, which finally outputs a desired phase-state. For an overview of how these protocols fit together the reader may refer back to Fig.~(\ref{fig_overview}).

\subsection{The parity-checker protocol} \label{secmsd}
 
Here we introduce a simple qutrit distillation protocol that is very efficient against a specific type of noise, but vulnerable against another type of noise.  However, both $\ket{H_{+}}$ and $\ket{\varphi}$ have zero overlap with the ``bad" noise term and so the protocol can be efficiently used to convert these states into a \textit{plus}-state.  Before beginning the iterative protocol some manipulation of the input states $\ket{H_{+}}$ and $\ket{\varphi}$ is required:
\begin{enumerate}
	\item	\textit{Preparation 1}, uniformly randomly choose from the set of unitaries $\{  1, H^{2} \}$ and apply;
	\item	\textit{Preparation 2},  apply $X^{\dagger} ;$ 
	\item	\textit{Preparation 3}, uniformly randomly choose from the set of unitaries $\{  1, S, S^{2} \} $and apply; 
\end{enumerate}
where  $S=  \kb{0}{0}+ \kb{1}{1}+ \omega \kb{2}{2}$.  For eigenstates of $H^{2}$ the first step is not strictly required, but it is listed to increase the generality of the protocol.
In particular, this preparation procedure maps all quantum states to
\begin{equation}
	\rho(\delta_{0}, \eta) = (1-\eta_{0} - \delta_{0}) \kb{\Psi^{+}}{\Psi^{+}} + \delta_{0} \kb{\Psi^{-}}{\Psi^{-}}  + \eta_{0} \kb{2}{2},
\end{equation}
where $\ket{\Psi^{\pm}} =( \ket{0} \pm \ket{1})/\sqrt{2}$.  Note that $\ket{\Psi^{-}} $ is Clifford equivalent to the Hadamard eigenstate $\ket{H_{i}}$.  For imperfect $\ket{H_{+}}$ and $\ket{\varphi}$ states, with depolarizing noise $\epsilon$, we have
\begin{eqnarray}
	\eta_{0} & = &  c(\ket{\psi}) + \hus{\epsilon} / 3, \\ 
	\delta_{0} & = & \hus{\epsilon} / 3,
\end{eqnarray}
where for the two magic states of interest $c(\ket{H_{+}})=0.2113$ and $c(\ket{\varphi})=0.0152$.  The parity-checker protocol will exponentially suppress the value of $\eta_{0}$, whereas $\delta_{0} $ will linearly increase.  However, this is not problematic as $\delta_{0} $ can be made arbitrarily small via distillation by the 5-qutrit protocol.

The iterative parity-checker is now fairly simple, on the $(n+1)$th round we have
\begin{enumerate}
	\item Take two copies of $\rho(\delta_{n}, \eta_{n})$;
	\item Measure the observable $Z_{1}Z_{2}^{\dagger}$ and postselect on +1;
	\item Decode the state such that $\ket{j,j} \rightarrow \ket{j}$;
	\item Use the output state $\rho(\delta_{n+1}, \eta_{n+1})$ as an input in the next iterate.
\end{enumerate}
It is straightforward to verify the iterative relations are
\begin{eqnarray}
	\eta_{n+1} & = & \eta_{n}^{2} /  p_{n}, \\ 
	\delta_{n+1} & = & \delta_{n} (1- \eta_{n}-\delta_{n})  / p_{n},
\end{eqnarray}
where $p_{n}$ is the success probability
\begin{equation}
	p_{n} = (1+\eta_{n}(3 \eta_{n} - 2))/2 .
\end{equation}
When the small noise component is zero, so $\delta_{0}=0$, and the large noise is not too large, $\eta_{0}< 1/3$, then $\eta_{n}$ vanishes exponentially quickly such that $\eta_{n} \sim (2 \eta_{0})^{n}$.  Allowing for non-zero $\delta_{0}$, the protocol can be iterated for approximately $n \sim \log (\delta_{0})$ rounds before the $\delta$ noise becomes problematic.  

Let us consider a concrete example. If we have a $\ket{\varphi}$ magic state with depolarization noise $\epsilon=10^{-8}$, this is first prepared into a noisy plus-state with $\eta_{0} \sim 0.0152$ and $\delta_{0} = 10^{-8}/3$.  The total noise, $\eta_{n}+\delta_{n}$, will decease for the first 3 rounds of parity checking.  After the 4th round we have a plus-state with a total error of only $2.707 \times 10^{-8}$.   This illustrates that high-fidelity plus-states can be prepared from high fidelity $\ket{\varphi}$ states in a small number of rounds.

\subsection{Equatorialization}
\label{secEquator}

Here we describe a simple magic state protocol that converts the plus-states to the phase-states. The phase-states lie on a generalization of the qubit Bloch sphere equator, hence the term \textit{Equatorialization}. This protocol is probabilistic but not iterative. We take two highly purified copies of a plus-state, $\ket{\Psi^{+}}$.  We measure a 2-qutrit stabilizer operator and postselect such that we project onto the subspace spanned by:
\begin{eqnarray*}
	\ket{0_{L}} & = & ( \ket{0,0} + \omega \ket{1,2} + \omega^{2} \ket{2,1} )  , \\ \nonumber
	\ket{1_{L}}= X_{1}X_{2}\ket{0_{L}} & = & ( \ket{1,1} + \omega \ket{2,0} + \omega^{2} \ket{0,2} )  , \\ \nonumber
	\ket{2_{L}}=X_{1}^{2}X_{2}^{2}\ket{0_{L}}  & = & ( \ket{2,2} + \omega \ket{0,1} + \omega^{2} \ket{1,0} )  ,
\end{eqnarray*}
and then decode onto a single qutrit. When successful this produces the following transformation:
\begin{equation}
	\ket{\Psi^{+}}^{\otimes 2} \rightarrow ( \ket{0} + \ket{1} + (\omega+\omega^{2}) \ket{2}) /\sqrt{3}.
\end{equation}
Noticing that $\omega+\omega^{2}=-1$, we find that the output is a phase-state
\begin{eqnarray}
	\ket{\Phi_{0, \pi}} =  ( \ket{0} + \ket{1} - \ket{2}) /\sqrt{3}.
\end{eqnarray} 
Finally, we have a magic state of the desired form.

\subsection{Gate injection}
\label{secGateInject}

Let us \hus{begin} by considering a general phase-state $\ket{\Phi_{\theta, \phi}}$. Given such a magic state, and a second qutrit state in any state, $\ket{\psi}$, we perform gate injection by measuring $Z_{1} Z_{2}^{\dagger}$.  Given measurement outcome $\omega^{k}$, we perform the decoding $\ket{x , y} \rightarrow \ket{ y + k}$.  The $\omega^{k}$ outcome effects a unitary, $U_{k,\theta, \phi}$, on $\ket{\psi}$, that is diagonal in the computational basis with eigenvalues
\begin{eqnarray}
	U_{0,\theta, \phi} & = & ( 1, e^{i \theta}, e^{i \phi}) , \\ \nonumber
	U_{1,\theta, \phi} & = & (  e^{i \theta}, 1, e^{i \theta}) , \\ \nonumber
	U_{2,\theta, \phi} & = & ( e^{i \phi}, e^{i \theta}, 1 ) .
\end{eqnarray}
Each unitary occurs with equal probability because the phase-state contains no variation in amplitudes.  Although random, a desired unitary can eventually be reached by repeated attempts, with each attempt corresponding to a step of a random walk on a manifold of phase gates with a toroidal topology.  For the $\ket{\Phi_{0, \pi}}$ state, the corresponding unitaries take a simple form.  The closure of $U_{k, 0, \pi}$ gives a group of order 4, \hus{up to} to a global phase, which is composed of $U_{k, 0, \pi}$ and the identity.  For such a small group any desired unitary will be reached, with high probability, within a small number of attempts.

For brevity, let us herein denote $N= U_{0, 0, \pi}$.  Clearly, $N$ is non-Clifford and so we label the group $\mathcal{C}_{+N}$ as the single qutrit group generated by $N$ and the Clifford group.  The  Gottesman-Knill theorem no longer applies and so we know no method of simulating computations using these unitaries.  However, does this provide a dense cover of SU(3)?  We discuss this question in App.~\ref{APPpromotion}, where we show that the group $\mathcal{C}_{+N}$ is of infinite order.  The size of the group, and that it contains basis changing gates of the Clifford group, make it highly plausible that it provides a dense cover of SU(3).  However, we presently do not have a complete proof.

\section{Summary}

In this paper we have demonstrated that  magic state distillation protocols do exist for higher dimensional systems. 
We have provided a generic formulation that can be used to study the distillation capabilities of any stabilizer code for any prime dimension. We have shown that the five qutrit code $[[5,1,3]]_3 $ is capable of distilling the qutrit Hadamard eigenstates, implying that, in analogy to the qubit case, there exist $ H$-type qutrit magic states. Under depolarizing noise the $ \ket{H}_{\pm} $ are distillable up to a noise threshold of $ 23.3 \% $. \hus{This is a higher threshold than the threshold achieved if the five qubit code were used to distil the qubit Hadamard eigenstates, which is $ \frac{1}{3}(3-\sqrt{6})\approx 18.3\% $ (see note \cite{Note1})}. We have introduced a seven qutrit code and shown how this code can improve the distillation resource region of $[[5,1,3]]_3 $. Interestingly, for the case of the seven qubit code, the distillation threshold of the Hadamard eigenstates is tight. And although the seven qutrit code does not \hus{distil} the Hadamard eigenstates, the numerical results of section A, suggests that this tightness is retained in the qutrit case, where all the mixed states along \hus{the non-stabilizer segments of the} the convex line of $ \ket{H_{\pm}} $ are distillable. An important difference with respect to the qubit case is the existence of one Hadamard eigenstate which is not an attractor to the distillation protocols studied, but which forms an unstable fixed point. This state is also the Hadamard eigenstate furthest from the stabilizer region. 

After searching the state space for distillable regions, we have found another set of distillable states outside the Hadamard plane. These states are eigenstates of a non-degenerate Clifford operator, and whether they are distillable by any other stabilizer code or are unique to the five qutrit code, is an open question. Under depolarizing noise, these states are distillable up to a noise threshold of $ 34.4\% $. 

We have also shown how to convert the outputs of the 5-\hus{qutrit} code into other magic states.  In turn, these have been shown to provide a unitary gate that promotes the Clifford group.  We have offered good, but inconclusive, evidence that this promoted Clifford group enables universal quantum computing. 

Finally, we list some few open questions that merit future investigation. 
Do analogous magic state distillation protocols work similarly in arbitrary dimensions (or maybe just prime dimensions)?
Are the magic states listed here exhaustive (could there be other attractive fixed points)?
Can we find a closed form for the $ H^{2} $ eigenstate coefficients?
Can we better understand the relationship between a code and the magic states associated with the code?
Is the set of unitaries (Clifford + (1,1,-1)) approximately universal?
Are there topological quantum computing models which realise all or part of the qudit Clifford group?

 
\section*{Acknowledgements}
We would like to thank Matty Hoban and Joe Fitzsimons for useful discussions, and also Chris Ferrie for his valuable comments on the manuscript. We acknowledge the financial support of the EPSRC, the Leverhulme Trust and the EU (QESSENCE).
\appendix

\section{Clifford equivalences and cycling behaviour}
\label{APPcycle}

\begin{figure}
    \includegraphics[height=7.5cm,width=7cm]{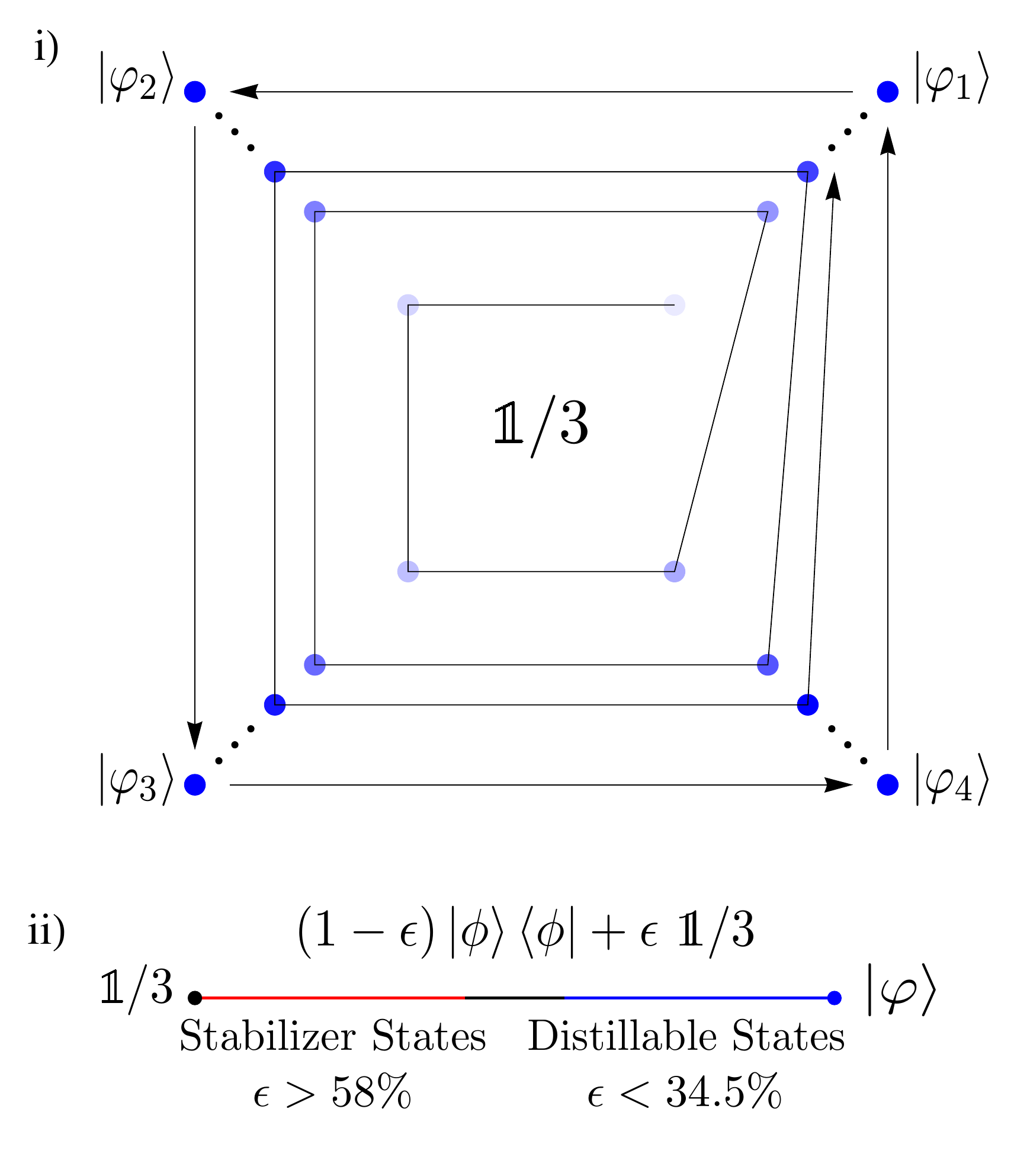}
    \captionsetup{singlelinecheck=yes, justification=centerlast}
      \caption{An illustrative picture of the cycling behaviour of $ \ket{\varphi} $. i) Starting with a mixed state (light blue point) the protocol will increase the purity of the states while cycling between them and ultimately reaching the fix pure points (dark blue points) . ii) The convex line between one of the cycling states $ \ket{\varphi}$ and the completely mixed state $ \openone/3 $ with an accurate ratio of the noise threshold.}
      \label{4cyc}
\end{figure}

We have chosen a particular decoding for which the distillation protocol has the simplest behaviour.  If instead, the canonical decoding was used, without the addition of an $R$ rotation, then purification would still occur, albeit between each iterate the output would cycle between different states.  In the Hadamard plane, we would observe an oscillation between $\ket{H_{\pm}}$, which is obvious since $R \ket{H_{\pm}} = \ket{H_{\mp}}$.  Whereas, for the $\ket{\varphi}$ state there is a more complex 4-cycle, illustrated in Fig.~\hus{(\ref{4cyc})}, such that for the distillation map for one iterate, $\mathcal{E}$, performs $\mathcal{E}(\ket{\varphi_{j}})=\ket{\varphi_{j+1}}$ and $\ket{\varphi}=\ket{\varphi_{1}}=\ket{\varphi_{5}}$.  The 4-cycling states are related by;
\begin{eqnarray}
	\ket{\varphi_{2}} & = & R^{\dagger} \ket{\varphi_{1}}, \notag\\ 
	\ket{\varphi_{3}} & = & H \ket{\varphi_{1}}, \\ \nonumber
	\ket{\varphi_{4}} & = & R^{\dagger}R^{\dagger} \ket{\varphi_{1}}.\notag
\end{eqnarray}
However, by considering $\mathcal{E'}(\rho)= R \mathcal{E'}(\rho) R^{\dagger}$, this cycling behaviour vanishes.  Note also, that this cycling behaviour is not only seen for the pure states but for depolarized states, and so all of these states are distilled by the 5-\hus{qutrit} code.

\section{$[[5,1,3]]_{3}$ Distillation}\label{appdistillation}

\begin{table}
\begin{center}
\begin{tabular}{|lccccc|}
\hline
$ g_{1} =$&$\sigma_{1,0}$&$\sigma_{0,1}$&$\sigma_{0,-1}$&$\sigma_{-1,0}$&$\sigma_{0,0}$\\
$ g_{2} =$&$\sigma_{0,0}$&$\sigma_{1,0}$&$\sigma_{0,1}$&$\sigma_{0,-1}$&$\sigma_{-1,0}$\\
$ g_{3} =$&$\sigma_{-1,0}$&$\sigma_{0,0}$&$\sigma_{1,0}$&$\sigma_{0,1}$&$\sigma_{0,-1}$\\
$ g_{4} =$&$\sigma_{0,-1}$&$\sigma_{-1,0}$&$\sigma_{0,0}$&$\sigma_{1,0}$&$\sigma_{0,1}$\\
\hline
$ \sigma_{1,0}^{L} =$&$\sigma_{0,1}$&$\sigma_{0,1}$&$\sigma_{0,1}$&$\sigma_{0,1}$&$\sigma_{0,1}$\\
$ \sigma_{0,1}^{L} =$&$\sigma_{1,0}$&$\sigma_{1,0}$&$\sigma_{1,0}$&$\sigma_{1,0}$&$\sigma_{1,0}$\\ \hline
\end{tabular}
    \captionsetup{singlelinecheck=yes, justification=centerlast}
	\caption{The stabilizer generators of the five qudit code \hus{and the qutrit logical operators} expressed in the $ \sigma_{j,k} $ notation.}
	\label{gensig}
\end{center}
\end{table}

Based on the generalized magic state protocol constructed in Sec. \ref{secmsd}, we will compute the exact distillation map on the Bloch components after a single round of distillation of the five qutrit code.

We start by expressing the stabilizer generators and the logical operators of the $[[5,1,3]]_{3}$ in terms of the  $ \sigma_{j,k} $ operators as shown in table \ref{gensig}. The input to the distillation protocol is $ 5 $ identical copies of a qutrit state $ \rho $. Using \hus{Eq.} (\ref{nrho}), the input state is expressed as:
\begin{equation}\label{rho5}
\rho^{\otimes 5}=\frac{1}{3^5}\displaystyle\sum_{(\boldsymbol j,\boldsymbol k)\in\mathbb{Z}_{3}^{5}}\alpha_{j_{1}\dots j_{5},k_{1}\dots k_{5}}\sigma_{j_{1}\dots j_{5},k_{1}\dots k_{5}},
\end{equation}
Using \hus{Eq.} (\ref{genproj}), a successful measurement of the  four stabilizer generators with outcome $ +1 $ corresponds to the following projector:
\begin{equation}
\Pi=\frac{1}{3^{4}}\displaystyle\sum_{\boldsymbol q\in\mathbb{Z}_{3}^{4}}g_{1}^{q_{1}}g_{2}^{q_{2}}g_{3}^{q_{3}}g_{4}^{q_{4}}.
\end{equation}
Substituting the stabilizer generators in table \ref{gensig} into the above expression and using the composition law in \hus{Eq.} (\ref{composition}), the projector becomes
\begin{align}\label{proj5}
\Pi=&\frac{1}{81}\displaystyle\sum_{\boldsymbol q\in\mathbb{Z}_{3}^{4}}\Bigl(\sigma_{(q_{1}-q_{3}),(-q_{4})}\otimes\sigma_{(q_{2}-q_{4}),(q_{1})}\otimes \sigma_{(q_{3}),(-q_{1}+q_{2})}\notag\\
&\hspace*{10mm}\otimes\sigma_{(-q_{1}+q_{4}),(-q_{2}+q_{3})}\otimes\sigma_{(-q_{2}),(-q_{3}+q_{4})}\Bigr).
\end{align}
We can simplify the notation of the above expression by using \hus{Eq.} (\ref{nfold}), which get rid of the tensor product sign, as shown in \hus{Eq.} (\ref{projsimp}).

The distillation map in \hus{Eq.} (\ref{genmap1}) can be put into a simpler form as follows:
\begin{align}
\alpha_{j,k}^{out}=&\frac{\mbox{tr}(\Pi\rho^{\otimes n}\Pi^{\dagger}(\sigma_{j,k}^{L})^{\dagger})}{\mbox{tr}(\rho^{\otimes n}\Pi)},\\
=&\frac{\mbox{tr}(\rho^{\otimes n}\Pi\sigma_{-j,-k}^{L})}{\mbox{tr}(\rho^{\otimes n}\Pi)}.\label{genmap}
\end{align}
where in the last step \hus{Eq.} (\ref{blochcomp}), $ \Pi =\Pi^{\dagger}$, $[\sigma_{j,k}^{L},\Pi] $=0 and the cyclic property of the trace were used.

The remaining task is to substitute \hus{Eqs.} (\ref{rho5}) and (\ref{proj5}) into \hus{Eq.} ({\ref{genmap}}) to calculate the distillation map on the Bloch components. 

Lets start by evaluating $ \mbox{tr}(\rho^{\otimes 5}\Pi) $. Recall that all the $ \sigma_{j,k} $ operators are traceless except for the identity operator $ \sigma_{0,0} $. Therefore, the only terms that will survive in $ \mbox{tr}(\rho^{\otimes 5}\Pi) $ are the coefficients of the identity operator. We get the identity operator in $ \rho^{\otimes 5}\Pi $ when the $ \sigma_{j,k} $ operators in $ \rho^{\otimes 5} $ and the $ \sigma_{j',k'} $ in $ \Pi $ have the opposite subscripts (i.e. $\sigma_{j,k}\sigma_{j',k'}=\sigma_{0,0} $ iff $ j=j' $ and $ k=k' $). As a result, $ \mbox{tr}(\rho^{\otimes 5}\Pi) $ will be the sum of all the Bloch components that are the coefficient of the $ \sigma_{j,k} $ operators such that the subscripts $ (j,k) $ are the negative of the subscripts in \hus{Eq.} (\ref{proj5}). In fact, since the summation is over all the elements of the ring $ \mathbb{Z}^{4}_{3} $ it is possible to multiply all the subscripts by $ (-1) $ without changing the actual value of the summation. Hence, $ \mbox{tr}(\rho^{\otimes 5}\Pi) $ can be compactly expressed as shown in \hus{Eq.} (\ref{trrhoproj}). In a similar way, we can express $ \mbox{tr}(\rho\Pi\sigma_{-j,-k}^{L}) $ for all four logical operators. For example, in the case of evaluating the output Bloch component $ \alpha_{1,0}^{\small{out}} $, \hus{Eq.} (\ref{genmap}) becomes
\begin{equation}
\alpha_{1,0}^{\small{out}}=\frac{\mbox{tr}(\rho^{\otimes 5}\Pi\sigma_{-1,0}^{L})}{\mbox{tr}(\rho^{\otimes 5}\Pi)},
\end{equation}
with $\mbox{tr}(\rho^{\otimes 5}\Pi\sigma_{-1,0}^{L})$ given in \hus{Eq.} (\ref{outbloch}).

We have evaluated the expressions for the four output Bloch components. However, writing them out in terms of $ \alpha_{j,k} $ notation is cumbersome. Therefore, for clarity, we will relabel the four qutrit Bloch components as follows $(\alpha_{1,0},\alpha_{0,1},\alpha_{1,1},\alpha_{1,2})\equiv(A,B,C,D)  $. For example, $ \mbox{tr}(\rho^{\otimes 5}\Pi) $ is given in \hus{Eq.} (\ref{dyneqns1}), where the subscript $ r $ represent the number of the distillation rounds with $ r=0 $ corresponding to the initial input state. Furthermore, it can be shown that the resultant expressions for the four output Bloch components can compactly be expressed in terms of a single function. This function is given in \hus{Eq.} (\ref{dyneqns2}), and based on this function the distillation map can be expressed as
\begin{align}\label{mapa}
A_{r+1}=&\mathcal{F}(A,B,C,D),\\
B_{r+1}=&\mathcal{F}(B^*, A, D, C^*),\\
C_{r+1}=&\mathcal{F}(A^*, C, B, D),\\
D_{r+1}=&\mathcal{F}(B^*, D^*, A^*, C).\label{mapd}
\end{align}

These four expression represents the complete distillation map, as the remaining four components are simply the complex conjugates of these expressions. However, the above expressions do not incorporate the additional corrective Clifford (see Sec. \ref{5qutritcode}), and will result to the cycling behaviour in App. \ref{APPcycle}. We need to ensure that the map in \hus{Eq.} (\ref{refR}) is applied after every iteration. This can easily be achieved in our formalism by the appropriate relabelling as follows: 
\begin{align}\label{mapa2}
A_{r+1}=&\mathcal{F}(D^{*},C,B^{*},A),\\
B_{r+1}=&\mathcal{F}(C^*, D^{*}, A, B),\\
C_{r+1}=&\mathcal{F}(D, B^*, C, A),\\
D_{r+1}=&\mathcal{F}(C^*, A^*, D, B^*),\label{mapd2}
\end{align}
which is the corrected distillation map. We can see that in order to calculate the fixed points of this map analytically, one would have to solve the above simultaneous complex multi-variable polynomials of order $ 5 $. It is known from the famous Abel-Ruffini theorem that there is no algebraic solution for a general polynomial of order five or above. Therefore, the best way to discover the fixed points of the distillation is through numerical means. We started with initial states $ \rho(A,B,C,D) $ for certain Bloch components and computed the above expressions for a number of iterations, and observed whether there is a convergence toward a fixed point. 

\begin{widetext}
\begin{align}
\Pi=&\frac{1}{81}\displaystyle\sum_{\boldsymbol{q}\in\mathbb{Z}_3^4}\sigma_{(q_{1}-q_{3})_{1}(q_{2}-q_{4})_{2}(q_{3})_{3}(-q_{1}+q_{4})_{4}(-q_{2})_{5},(-q_{4})_{1}(q_{1})_{2}(-q_{1}+q_{2})_{3}(-q_{2}+q_{3})_{4}(-q_{3}+q_{4})_{5}}\label{projsimp}.\\
\mbox{tr}(\rho^{\otimes 5}\Pi)=&\frac{1}{81}\displaystyle\sum_{\boldsymbol{q}\in\mathbb{Z}_3^4}\alpha_{(q_{1}-q_{3})_{1}(q_{2}-q_{4})_{2}(q_{3})_{3}(-q_{1}+q_{4})_{4}(-q_{2})_{5},(-q_{4})_{1}(q_{1})_{2}(-q_{1}+q_{2})_{3}(-q_{2}+q_{3})_{4}(-q_{3}+q_{4})_{5}}\label{trrhoproj}.\\
\mbox{tr}(\rho^{\otimes 5}\Pi\sigma_{-1,0}^{L})=&\frac{1}{81}\displaystyle\sum_{\boldsymbol{q}\in\mathbb{Z}_3^4}\alpha_{(q_{1}-q_{3})_{1}(q_{2}-q_{4})_{2}(q_{3})_{3}(-q_{1}+q_{4})_{4}(-q_{2})_{5},(-q_{4}+1)_{1}(q_{1}+1)_{2}(-q_{1}+q_{2}+1)_{3}(-q_{2}+q_{3}+1)_{4}(-q_{3}+q_{4}+1)_{5}}\label{outbloch}. 
\\
\label{dyneqns1}
  \mbox{tr}\left(\rho^{\otimes 5}\Pi\right)=& \frac{1}{81}\bigl( 1+ 10 \left(\left|A_r\right|{}^2+\left|D_r\right|{}^2\right) \left(\left|B_r\right|{}^2+\left|C_r\right|{}^2\right)+5\bigl(B_r^2 A_r^* C_r^{*2}+ D_r^2
   A_r^{*2} B_r^*+ \notag\\
  & D_r \left(A_r^2 D_r C_r^*+B_r^2 C_r^2\right)+ B_r^{*2} \left(A_r C_r^2+C_r^{*2} D_r^{*}\right)+D_r^{*2} \left(A_r^2 B_r+C_r A_r^{*2}\right) \bigr)\bigr).
\\
\mathcal{F}(A,B,C,D)=&\frac{1}{81}\Bigl(B_r^5 +
10  B_r \left(D_r A_r^*+B_r^*\right) \left(A_r C_r^*+ C_r D_r^*\right)+ 
5 \bigl(A_r C_r^2 \left|A_r\right|^2+D_r^2 \left(A_r B_{r}^{*2}+C_r\right)+\notag\\
   &A_r^{*2} \left(B_r^{*2}D_r^*+C_r^*\right)+D_r^{*2} \left(A_r^2+D_r
   C_r^{*2}\right)+\left|C_r\right|^4 B_r^*\bigr)\Bigr)/ \mbox{tr}\left(\rho^{\otimes 5}\Pi\right).
\label{dyneqns2}
\end{align}
\end{widetext}

\section{The size of the promoted Clifford group}
\label{APPpromotion}

Here we show that $\mathcal{C}_{+N}$ contains an infinite number of unitaries.  We do so by considering a particular unitary,
\begin{equation}
	K = H.N.H.N.H ,
\end{equation}
which is an alternating sequence of Hadamards and our non-Clifford gate.  The eigenvalues, $e^{i \pi \lambda_{n}}$ of $K$ can be analytically calculated by Mathematica, such that
\begin{eqnarray}
	\lambda_{1} & = & \mathrm{ArcTan} ( \sqrt{2}) / \pi , \\ 
	\lambda_{2} & = & -\mathrm{ArcTan} ( \sqrt{2}) / \pi , \\ 
	\lambda_{3} & = &  1 / 2. 
\end{eqnarray}
Since both $\lambda_{1}$ and $\lambda_{2}$ are irrational numbers, every power of $K$ gives a different unitary, and so the single qutrit group, $\mathcal{C}_{+N}$, is infinite in order. The irrationality of $\lambda_{1}$ , and hence also $\lambda_{2}$, can be shown by expressing it as $\lambda_{1}= \mathrm{ArcCos}( 1 /\sqrt{3}) / \pi$ and then using a theorem from \cite{Book09}.

\end{document}